# *Negative selection maintains transcription factor binding motifs in human cancer*

Ilya E. Vorontsov [2], Ivan V. Kulakovskiy [1,2]*, Grigory Khimulya [2],

Elena N. Lukianova [4], Daria D. Nikolaeva [5], Irina A. Eliseeva [6],

and Vsevolod J. Makeev [2,1,3]

[1] Engelhardt Institute of Molecular Biology, 119991, GSP-1, Vavilova 32, Moscow, Russia

[2] Vavilov Institute of General Genetics, Russian Academy of Sciences, 119991, GSP-1, Gubkina 3, Moscow, Russia

[3] Moscow Institute of Physics and Technology State University, 141700, Dolgoprudny, Moscow Region, Russia

[4] Center of Genetics and Reproductive Medicine, Human Stem Cells Institute, 119991, Gubkina 3-1, Moscow, Russia

[5] Lomonosov Moscow State University, Faculty of Bioengineering and Bioinformatics, 119991, GSP-1, Vorobyevy Gory 1-73, Moscow, Russia

[6] Institute of Protein Research, Group of Protein Biosynthesis Regulation, 142290, Institutskaya 4, Pushchino, Russia

*corresponding authors

ivan-dot-kulakovskiy-at-gmail-dot-com

vsevolod-dot-makeev-at-gmail-dot-com

*Abstract*

**Background**

Somatic mutations in cancer cells affect various genomic elements disrupting important cell functions. In particular, mutations in DNA binding sites recognized by transcription factors can alter regulator binding affinities and expression of target genes. A number of promoter mutations have been linked with an increased risk of cancer, mutations in binding sites of selected transcription factors have been found under positive selection. However, negative selection of mutations in coding regions is elusive and significance of negative selection in non-coding regions remains controversial.

**Results**

Here we present analysis of transcription factors with binding sites co-localized with non-coding variants. To avoid statistical bias we account for mutation signatures of different cancer types. For many transcription factors, including multiple members of FOX, HOX, and NR families, we show that human cancers accumulate fewer mutations than expected by chance that increase or decrease affinity of binding motifs. Such conservation of motifs is even more exhibited in DNase accessible regions.

**Conclusions**

Our data demonstrate negative selection against binding sites alterations and suggest that this selection pressure protects cancer cells from rewiring of regulatory circuits. Further analysis of transcription factors and the respective conserved binding motifs can reveal cell regulatory pathways crucial for the survivability of various human cancers.

*Background*

Somatic mutations in DNA binding sites recognized by transcription factors [1, 2] can alter regulator binding affinities and expression of the target genes [3], often leading to malignant cell transformation. Affinity change can be directly associated with cancer progression with a striking example of a GABP (ETS-family factor) binding site emerging in TERT promoter [4] associated with progression of different tumor types [5, 6]. A number of other promoter mutations have been linked with an increased risk of cancer [7, 8], and the number is expected to grow rapidly with extensive sequencing of complete cancer genomes. For instance, in the recent study of regions associated with a risk of epithelial ovarian cancer [9] out of nearly three hundreds significant SNPs only two were found in protein coding regions, whereas 25 such SNPs were localized in transcription factor binding sites.

Likewise, cancer drivers identified in knockdown experiments [10] not necessarily carry mutations in coding regions, thus underlining the importance of regulatory mutations modifying gene expression.

Frequencies of synonymous and non-synonymous substitutions allow studying selection of cancer somatic mutations in protein-coding regions [11]. For non-coding regions estimates of selection pressure can be based on functional annotation of sequence variants. In particular, DNA sequence motifs [12] recognized by the transcription factors usually have strict and degenerate positions. This allows assessing selection of sequence variants in binding sites in a way resembling usage of non-synonymous and synonymous substitutions in codons.

Somatic mutations often tend to destroy binding sites [3, 13] of specific transcription factors reflecting positive selection of variants. Conversely, binding sites of other transcription factors were reported to avoid mutational changes [13], but the significance of negative selection pressure acting at somatic mutations remains controversial [3, 13].

Mutations in cancer cell lineages are strongly context-dependent [14]; thus, mutation signatures of different cancer types should be properly taken into account to avoid statistical bias. Here, we used genome wide data [14] on several cancer types with different mutation signatures to study the frequencies of somatic mutations that alter binding sites for specific transcription factors.

## Results

*Assessing selection pressure on transcription factor binding sites*

To study selection pressure on gene regulatory regions we used mutation sets from different cancer samples grouped by the tissue [14]. First, we selected mutations in putative regulatory regions (intronic and promoter genomic segments) which made up to 50% of total mutation calls (Additional file 3: Supplementary Table 1). Then, we mapped transcription factor binding sites [15] in small windows centered at the mutation sites (see Methods, predicted binding sites were allowed to be located in the vicinity of but not necessarily overlap the mutated base). For a particular tested motif, on average 5 ± 3% of tested windows included binding sites predictions, but there were exceptional cases with notably deviating prediction rates (see Additional file 4: Supplementary Table 2 for complete data).

Putative affinity changes were estimated for the mutated allele versus the germline [16]. We separately considered both directions of affinity change that can be caused by a nucleotide substitution in a binding site: affinity loss (disruption of a binding site predicted for the germline allele) and affinity gain (improvement or emergence of a binding site with stronger prediction for the mutated allele).

To evaluate the selection pressure we compared the observed frequency of mutations substantially changing the binding site affinity with the expected frequency estimated from simulated control data.

We used two different control data sets: (1) the *shuffle* control consisting of sequences with randomly shuffled nucleotides around the actual mutated bases, similar to that in previous studies [3] but controlling the mutation context (the germline and mutated nucleotides and the proximal 5' and 3' nucleotides); and (2) the *genomic* control consisting of randomly sampled segments of promoter and intronic regions not overlapping the mutation-centered windows (see Methods for details).

To account for specific mutation signatures of different cancer types (see Additional file 1: Supplementary Figure 1), binding sites predictions in both shuffle and genomic controls were sampled to equalize the resulting distribution of mutation contexts of a given control data to match the cancer mutations data, separately for each cancer type.

Finally, we identified binding motifs exhibiting an exceptional rate of mutation-induced affinity changes versus both control data sets with equalized contexts distribution (FDR-corrected two-tail Fisher's P < 0.05, see Additional files 4-5: Supplementary Table 2-3 for complete data, see Methods for details).

*Limited magnitude of selection pressure requires high statistical power*

For each transcription factor binding motif we estimated magnitude of selection pressure on somatic mutations overlapping the predicted binding sites. The selection pressure magnitude was defined as the ratio of the observed and expected frequencies of mutation-induced affinity changes assessed for the somatic mutations (observed) and the simulated control data (expected), respectively (see Methods). The typical values of the selection pressure magnitude were around 0.9-0.95 (negative selection) and 1.05-1.1 (positive selection, see Figure 1) and were in a similar range for mutations causing affinity gain and affinity loss.

With ratios expressing selection pressure magnitude so close to 1, a large data volume was necessary to attain acceptable statistical significance. In particular, the simulated control sets were several times larger than the initial cancer data set, especially for cancer types with less called mutations. The most robust observations were made on cancer types with the highest mutation counts and thousands to dozens of thousands predictions per binding motif (see Additional file 4: Supplementary Table 2).

*Negative selection of mutations altering binding motif affinity*

Among transcription factors with binding sites experiencing frequent affinity loss we observed those belonging to AP-2 and C/EBP families, whose binding motifs were previously reported as mutation-enriched [3, 13]. Binding motifs of zinc finger SP and KLF families were also enriched with affinity loss-causing mutations in several cancer types. Mutations in binding sites of other transcription factors, in particular, ETS family motifs, persistently induced affinity gain (see Table 1 and Additional file 4-5: Supplementary Table 2-3). Binding motifs enriched with mutations causing affinity loss or affinity gain are likely to be under positive selection.

Conversely, for a much wider set of transcription factor binding motifs, mutations leading to either affinity gain or affinity loss were depleted (Table 1, Additional file 4-5: Supplementary Table 2,3). Furthermore, for some factors binding motifs were simultaneously protected from both affinity loss and gain in several cancer types. In particular, there were several families of nuclear receptors (Figure 2, TFClass families [17] are shown). Such conserved binding motifs indicate action of negative selection against somatic mutations. Importantly, negative selection of HOX and FOX motif-changing variants, that was reported earlier for normal tissues [18], was also exhibited for mutations in different cancers.

Only a few binding motifs were found significant for cancer types with limited number of available mutation calls due to lower statistical power. However, these orphan motifs often belonged to the families found under systematic protection from the affinity loss or gain in larger data sets (Additional file 5: Supplementary Table 3).

*Stronger negative selection acts in DNase accessible regions*

Accuracy of binding sites prediction *in silico* is limited and it is hard to distinguish true binding sites from false positive predictions without direct experimental data. To increase the confidence of binding site prediction, we considered subsets of mutations occurring in DNase accessible segments [19] of promoters and introns for breast cancer and lung adenocarcinoma.

Mutation rates can unpredictably depend on chromatin accessibility. Hence, a separate control set constructed from DNase accessible regions was necessary to evaluate selection of mutations in DNase accessible regions. The resulting estimates of the selection pressure magnitude became comparable with those for the whole set of mutations in promoter and intronic segments.

A smaller absolute number of mutations in DNase accessible regions resulted in a lower number of binding sites predictions and a lower statistical power (Additional file 6: Supplementary Table 4), thus the absolute number of featured binding motifs was also smaller. However, the major observations persisted. In particular, motifs of FOX and several NR families were found protected from somatic mutations whereas selected members of AP-2 and C/EBP families displayed persistent affinity loss.

Taking the motifs found under significant negative selection for the full set of intronic+promoter mutations ($P$-value < 0.05 for a particular control), we compared the estimates of the selection pressure magnitude with those for mutations in the DNase accessible regions. While there were no systematic difference for the shuffle control, the genomic control revealed consistently lower relative frequency of the affinity changing events (i.e. stronger selection magnitude) for the most of significant motifs in DNase accessible regions (Figure 3). We believe this is a strong indication of the increased negative selection pressure on DNase accessible regions. The magnitude of positive selection in DNase accessible regions was lower (closer to 1), and the number of motifs detected under positive selection was lower (Additional file 6: Supplementary Table 4).

## *Discussion*

### *Similar binding motifs are under similar selection pressure*

Transcription factors of a given structural family [17] usually share similar binding preferences and it is not always possible to distinguish binding sites bound by different members of the same family. In particular, attribution of binding predictions to a particular transcription factor is not entirely reliable, that is why focused on observations that were consistent for different members of a given motif family. Furthermore, the similarity of binding motifs of the same family made it less likely that the observed statistical preferences of mutations to alter or avoid predicted binding sites appeared from biased predictions of a particular low-quality binding model. Transcription factors of the same family often had binding motifs obtained from different experimental data sets [15] and thus had different prediction biases. Consistency between several motifs belonging to transcription factors of the same family increased our confidence that the detected selection pressure was indeed related to binding sites.

### *Genomic control data highlight negative selection*

We emphasize the usage of genomic control in addition to shuffle control, since genomic sequences prefer or avoid occurrences of sequence motifs in a non-random fashion. For instance, composition of CpG islands correlates with the presence of many Kruppel-like transcription factor motifs, whereas nucleosome binding motifs facilitate binding of TBP factor [20]. All these regularities are destroyed in shuffle control.

In general, genomic control gave more conservative estimates of selection magnitude (Figure 1) but there was a notable overlap between the resulting sets of motifs (Additional file 2: Supplementary Figure 2) identified in comparison with any of the two controls, especially for the motifs conserved by negative selection. However, for positive selection the overlap was quite limited. For example, binding sites of HIF-1 transcription factor were found under strong positive selection both for affinity gain and affinity loss when the shuffle control was used, but the effect completely disappeared with the genomic control (Additional file 4: Supplementary Table 2). Such observations are not easy to interpret, so in our analysis we focused on cases, which were consistent (with the same selection direction) and significant (P < 0.05) for both control data sets.

Many transcription factors with binding sites under selection have been reported to be involved in cell malignant transformation. For instance, we detected significant enrichment of affinity gain events for C/EBP that was reported to be important for malignant conversion of human breast epithelial cells [21]. Binding sites of GABP, a member of ETS family, are created by mutations in TERT promoter and associated with development of many cancer types [4]. In our study the affinity gain of ETS binding sites appeared under positive selection, whereas affinity loss under negative selection. FOX proteins, whose binding sites were found under negative selection both for affinity gain and loss, were also suggested to be involved in cancer progression [22].

To summarize our findings, we observed transcription factor binding sites of many motif families in several cancer types altered by somatic mutations significantly less frequently than expected, both for mutations causing affinity loss or gain. The avoidance of mutations in binding motifs indicated the action of negative selection maintaining specific paths in cellular regulatory circuits. This observed negative selection of mutations leading to substantial affinity gain rejects a possible alternative explanation that the observed difference is caused by transcription factors providing the protection against mutations at binding sites by occupying respective DNA segments. Another alternative explanation, that the observed statistical phenomena arise from biased mutation patterns of a particular cancer type, can be ruled out because conserved binding motifs are shared in cancers with substantially different mutation signatures (see Additional file 3: Supplementary Table 1 and Additional file 1: Supplementary Figure 1).

Finally, the conservation of binding motifs against mutations was exhibited even in a simple test considering how often mutations occupied positions within the motifs versus nearby positions in the vicinity (see Methods and Additional file 7: Supplementary Table 5). The resulting list of motifs was less selective and harder to interpret since the substitution itself and affinity change direction were not considered. However, major families of motifs that tended to overlap/avoid mutations (i.e. with more/less mutations within motifs than expected) were consistent with the detailed affinity change test, in particular, including members of ETS, AP-2 and FOX, HOX families overlapping and avoiding mutations, respectively.

Further analysis of factors with binding motifs protected by negative selection against mutations can reveal cell regulatory pathways crucial for the survivability of various human cancers.

## Methods

*Overview of cancer mutations data*

We used published whole genome somatic mutations data for ten cancer types [14] (507 samples with varying sequencing depth) with mutations from different samples of the same cancer type aggregated. The total number of mutation calls varied between cancer types with breast cancer, liver cancer and lung adenocarcinoma having the largest numbers (Additional file 3: Supplementary Table 1). Only single-nucleotide substitutions were considered.

Ensembl gene annotation was used to select mutations in [-5000;+500] bp intervals from transcription start sites (promoters) and intronic regions. The length of considered intronic and promoter segments totaled $1,6·10^9$ bps with an average mutations density of 1-2 substitutions per 5 kb for the cancers with the highest number of mutation calls.

Mutations in coding regions were excluded. An overview of the initial mutation data is given in Additional file 3: Supplementary Table 1. The relative frequencies of mutation contexts (the 5' and 3' nucleotides surrounding the mutated/germline alleles) are shown in Additional file 1: Supplementary Figure 1.

The mutation coordinates were used to extract mutation-centered [-50; +50] bp genomic windows based on Ensembl GRCh37.p13 (release 75) genome assembly. Overlapping windows for closely located mutations were considered independently (thus a single binding site in theory could be assessed for more than one mutation). Within a single cancer sample some windows with different mutation coordinates had identical sequences (e.g. due to genomic repeats), and, for each particular sample, only one of these windows was kept for further analysis. Recurrent mutations in different samples were considered as independent observations. The statistical analysis (see below) was performed separately for each cancer type.

*Assessing binding motif affinity changes*

DNA motifs recognized by transcription factors are highly divergent. A basic binding site model, position weight matrix (PWM), accounts for such divergence by assigning a score for each oligonucleotide of some fixed length, with high scoring sequences selected as binding sites. A fixed score threshold defines the positive prediction rate (motif *P*-value) for the given PWM. With a uniform distribution of background frequencies the motif *P*-value is equal to the fraction of oligonucleotides ("words") scoring above the given threshold among all words of the fixed length. For a given sequence variant the score of the best binding site prediction defines the respective motif *P*-value. In annotation of regulatory SNPs the ratio of motif *P*-values for two sequence variants was used to quantify the effects of nucleotide substitutions in predicted binding sites [23, 24].

The correspondence between the motif *P*-value and the energy of specific binding (the so-called discriminative energy) is clarified in the classic work of [25]. It is shown that for a point substitution in a binding motif the log-ratio of *P*-values defined by two alternative sequence variants is approximately equal to the difference of corresponding discriminative energies. We defined substantial difference as the drop in the discriminative energy that increases the motif *P*-value four fold, which corresponds to the substitution of a perfectly nondegenerate position to a completely degenerate position in an imaginary model that did not contain any degenerate positions.

To predict the binding sites and to quantify affinity change events we used PERFECTOS-APE software [16]. Binding sites were predicted in [-50;+50] bp mutation-centered windows as best hits of 278 A/B/C-quality (highest quality) PWMs from the HOCOMOCO [15] collection. The closest position of a binding site was required to be located not farther than 11 bp (one helix pitch) away from the mutated base. This setup allowed us to bypass global variability in mutational rates which may induce an unknown bias in co-localization of mutations with binding motifs. To avoid such conditioning by genome location preferences we predicted binding sites only in small windows centered at mutations, thus considering only regions containing both a binding site and a somatic mutation.

For predictions we used the motif *P*-value threshold of 0.0005 (which roughly defines a false positive rate with a single expected prediction per 1000bp of random double stranded sequence). The approach to take PWM score thresholds according to a common false positive rate was recently demonstrated to be the least biased in a comparative study [26].

It is not trivial to truly assess the quality of binding site prediction since very little is known on the negative control, the DNA sequences that do not bind a particular transcription factor. For example, FoxA2 has well-exhibited binding preferences and has been tested for false-positive predictions in [27] with the help of EMSA experiments. For the HOCOMOCO model at 0.0005 P-value the resulting experimentally-justified FDR was about 17% providing an intuition on binding site prediction error rate. Yet, since only 64 hand-picked binding sites (41 positive and 23 negative) were checked with EMSA this FDR value cannot be used in any quantitative estimation.

We used the *P*-value ratio thresholds of 4 (affinity gain, motif emergence) and 0.25 (affinity loss, motif disruption) to distinguish between substantial and non-substantial binding site alterations. This setup was identical for each particular cancer type (to estimate the observed affinity change frequencies) and control data (to estimate expected frequencies).

To sum up, for a given transcription factor binding motif for a given mutation-centered window an affinity loss event was counted if (1) the best prediction for the germline sequence passed motif *P*-value of 0.0005 considering PWM hits not farther than 11bp away from the mutated base and (2) the best prediction for the mutated allele, again, considering PWM hits not farther than 11bp away from the mutated base, had *P*-value at least 4 times weaker. Symmetrically, an affinity gain event was counted under the same restrictions for the best PWM hit but for the sequence with the mutated allele, and with the respective *P*-value for the germline sequence predictions being 4 times weaker.

We did not require the best hits in the germline and the mutant sequence to appear at the same position. Thus, with our approach we did not counted affinity change events in windows with two good motif hits, only one of which was affected by the mutation.

### *Simulated control data*

Two simulated control sequence sets, the shuffle control and genomic control, were used to estimate the expected frequency of affinity changing substitutions. The relative size of the simulated control sets depended on the total number of mutation calls in a particular cancer type. Lower relative size of the control data sets was used for the larger mutation sets (see Additional file 3: Supplementary Table 1). Higher relative size of the control data sets was used for the cancers with lower numbers of mutation calls to provide stable estimates of expected frequencies.

The shuffle control set was obtained by shuffling the flanking sequences within [-50;+50] bp around the mutated base keeping the mutation context, the immediate 5' and 3' nucleotides, and the substitution itself, intact. Multiple shuffles were gathered for each mutation (Additional file 3: Supplementary Table 1). This was the only step where the window length was explicitly used.

The windows for the genomic control were sampled from intronic and promoter regions in a way that they did not overlap the cancer mutation-centered windows. Each segment of [-50;+50] bp had the central base and its neighboring 5' and 3' nucleotides identical to the mutation context of a given somatic mutation locus, the respective nucleotide alternative was added. For each somatic mutation several genomic control windows were extracted, the number depended on the total number of mutations for a particular cancer type (Additional file 3: Supplementary Table 1).

Both the shuffle and genomic controls were used to predict transcription factor binding sites in the same way as for the cancer data. For each binding motif the windows with binding sites predictions for the germline alleles were used to evaluate statistical significance of the affinity loss. Likewise, the windows with binding sites predicted for the mutated alleles were used to evaluate statistical significance of the affinity gain. The windows with predictions for both alleles participated in both types of analysis (Figure 4), and the windows without predictions were discarded.

Since binding sites predictions depended on the nucleotide composition and, consequently, on the mutation contexts (the 5' and 3' nucleotides proximal to the mutated base), we equalized the mutation contexts distributions of the test and control data for each particular cancer type before the statistical evaluation (see below). To achieve this, we sampled the windows with binding sites predictions in control data (both shuffle and genomic) to match a given cancer data for each binding motif separately.

In a limited number of cases there were not enough control data to completely equalize the contexts distribution (see Additional file 4: Supplementary Table 2). Yet, even for cancer types with low number of mutation calls, where the relative required size of the control data sets was extremely large, no less than 95% of predictions with matching contexts were successfully sampled from the control data. Importantly, for cancer types with abundant mutation calls context equalization was almost perfect (99.9-100% match of the contexts distributions with the non-perfect match only for exceptional motifs, see Additional file 4: Supplementary Table 2), since a lower relative size of the control data set was generally required (see Additional file 3: Supplementary Table 1). During significance evaluation (see below) the "missing" control predictions were considered as those making the contingency tables more uniform (i.e. reducing the difference and its possible statistical significance).

Thus, for each binding motif we obtained the final sets of mutation-centered windows with binding sites overlapping with or located in the close vicinity of mutations for test and control data with an equalized mutation contexts distribution. This eliminated possible bias from the non-randomness of mutational signatures and made possible a comparison of the binding sites alteration frequencies in cancer versus control data.

The events of mutation-induced motif changes were counted for each cancer type and the control data sets (shuffle and genomic) using the same procedure. For each binding motif the Fisher's exact test was computed using 2x2 contingency tables (substantial affinity loss or gain versus non-substantial affinity change/no change, cancer mutations versus the control data, separately for shuffle and genomic control), refer to Figure 4 for a scheme.

Only cases that passed 0.05 FDR-corrected (for 278 tested binding motifs) Fisher's exact test *P*-value in both comparisons (versus the shuffle and versus genomic controls) were considered significant for a particular cancer type.

*Estimating the selection magnitude*

The affinity loss frequency was calculated as the fraction of affinity loss events out of all tested windows with legitimate motif predictions for the germline allele (see the previous section). Symmetrically, the affinity gain frequency was calculated as the fraction of with affinity gain events among all windows with legitimate predictions for the mutated allele.

The absolute values of the affinity loss and gain event frequencies are biased by the specific mutation signature of each particular cancer type. To account for this effect, we computed the relative values normalizing cancer frequencies for those of the control data. The ratio of the affinity loss (gain) frequency for cancer somatic mutations to the affinity loss (gain) frequency in the control set (genomic or shuffle) defined the selection magnitude for affinity loss (gain). The magnitude greater than 1 corresponded to the positive selection, the magnitude less than 1 corresponded to the negative selection.

### Assessing mutations in DNase accessible segments

As an additional test, we considered DNase accessible segments of all introns and promoters. The DNase accessibility data (breast cancer and lung adenocarcinoma only) were collected for related cell lines and normal tissues [28] (Additional file 6: Supplementary Table 4). The resulting reduced set consisted of 104905 (596253) mutations for breast cancer (lung adenocarcinoma) thus including nearly 30% (90%) from the respective total sets of intronic and promoter mutations.

Open chromatin regions are enriched with binding sites of the most of transcription factors but depleted of others [19]. To account for this non-randomness, the shuffle and the genomic controls were produced with the same pipeline as for the total set of promoter+intronic mutations but restricted to DNase accessible regions only.

### A simplified test to reveal selection pressure

We also tested a basic overlap of mutations and motifs not taking into account the affinity change. To this end we used 2x2 contingency tables for the test / control data and mutations overlapping / not overlapping with the motifs (Additional file 7: Supplementary Table 5) with the binding sites predictions performed using the same setup as in the main workflow. On the one hand, this simplified test did not capture neutral substitutions within the motifs and did not allow separating affinity loss and gain events. On the other hand, the results of this test did not depend on arbitrary selected motif *P*-value ratio thresholds.

## List of abbreviations

PWM: position weight matrix

## Competing interests

*The authors declare no competing financial interests.*

## Authors' contributions

IEV, IVK, and VJM designed the analysis pipeline and wrote the manuscript. GK, ENL, DDN, and IAE participated in the data analysis and manuscript preparation.

## Acknowledgements

The authors thank Andrei Zinoviev, Shamil Sunyaev, Mikhail Gelfand, Alexander Kel, Yulia Medvedeva and, specially, Fyodor Kondrashov for fruitful discussions. This work started during the School for Molecular and Theoretical Biology (Pushchino) organized by the Dynasty Foundation. We personally thank Dmitry Zimin and Anya Piotrovskaya.

This study was supported by RFBR grant 15-34-20423 and, partly, by RFBR grant 14-04-01838. Improvement of used motif analysis software was supported by Russian Science Foundation grant 14-50-00060. IVK was personally supported by the Dynasty Foundation Fellowship.

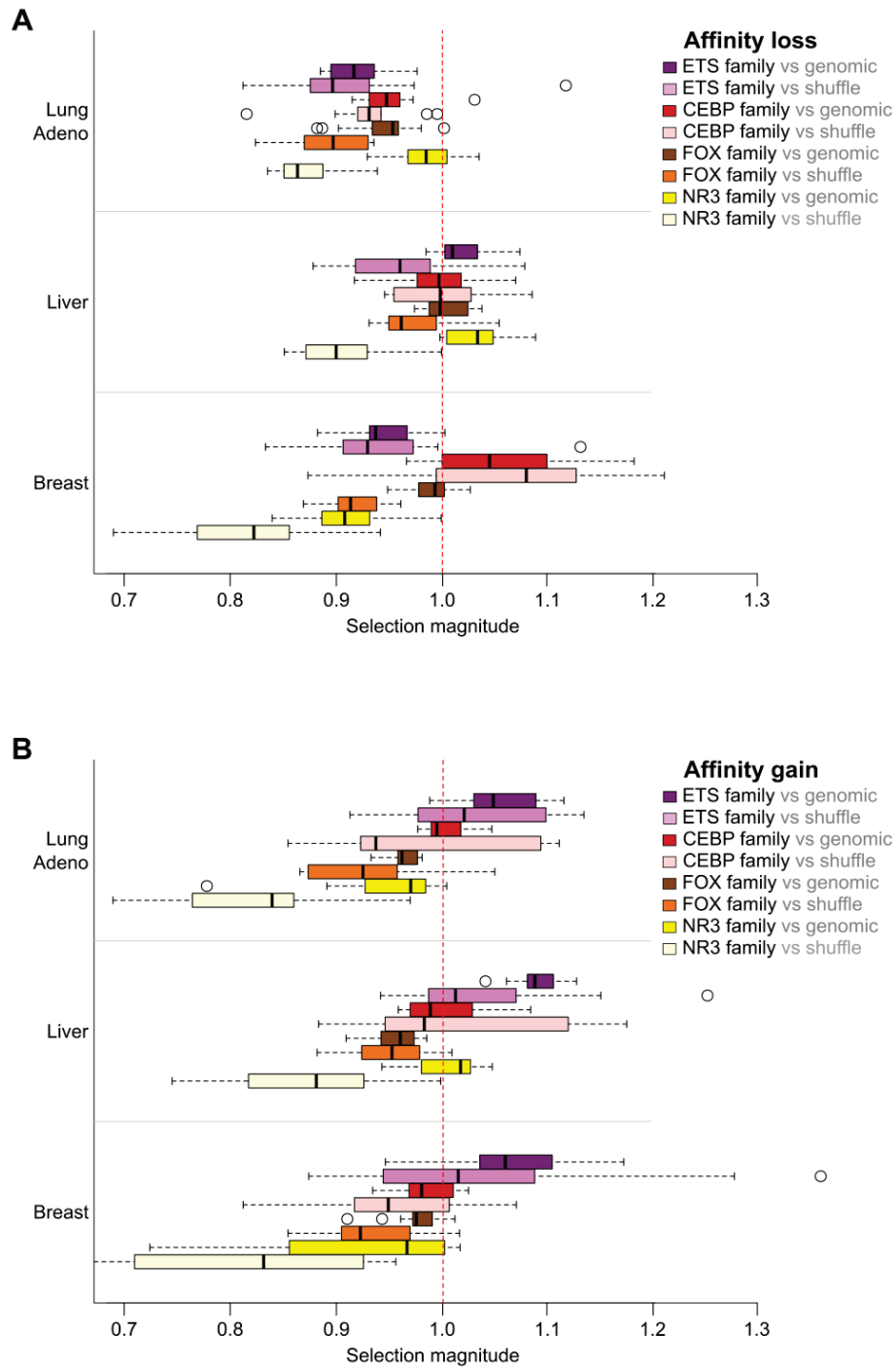

**Figure 1.** *Selection magnitude for affinity loss and gain of ETS, FOX and C/EBP motifs in different cancer types.*

X-axis displays the selection magnitude for motif affinity loss (A) or gain (B) caused by somatic mutations. Box-plots are provided for ETS-related (14 motifs), FOX (13 motifs), C/EBP-related (9 motifs) and NR3 (Steroid hormone receptors, 11 motifs) transcription factor families in three cancer types with the largest numbers of mutation calls. In particular, C/EBP motifs display frequent affinity loss in breast cancer, FOX and NR3 motifs are protected from both the affinity loss and gain in lung adenocarcinoma and breast cancer, and ETS motifs tend to emerge in all three cancer types (breast, lung and liver). Data for two control datasets (shuffle, genomic) are shown.

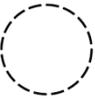

**Figure 2.** *Transcription factor binding motifs protected from somatic mutations in different cancer types.*

The size of a pie chart shows the total number of motifs in a given transcription factor family (given in curly braces according to TFClass). The slices of a pie show the number of conserved binding motifs protected from any affinity change (yellow), motifs protected from affinity loss (magenta), and motifs protected from affinity gain (deep purple).

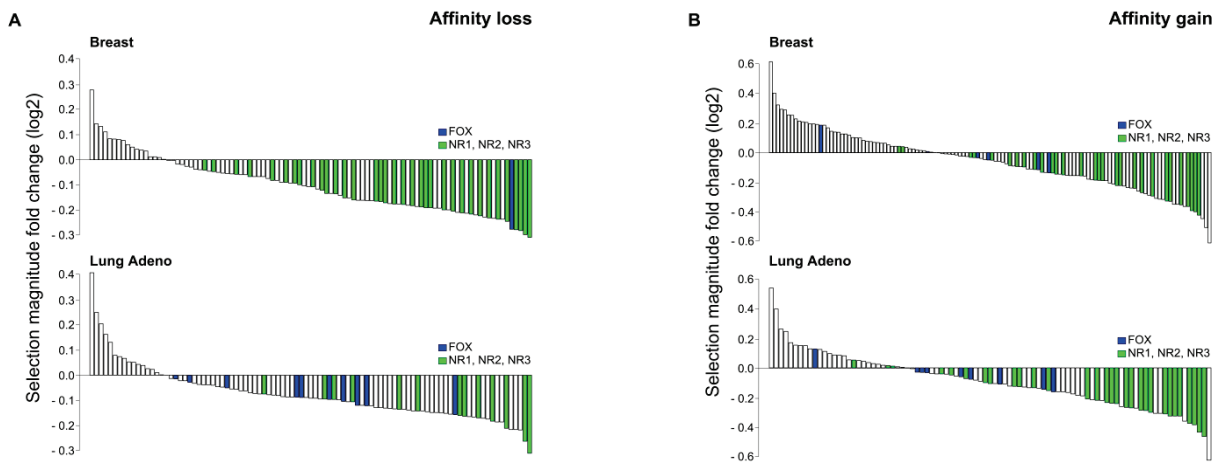

**Figure 3.** *Fold change (log2) of negative selection magnitude for mutations in DNase accessible subregions compared to that in the promoter and intronic segments.*

Y axis displays selection magnitude fold change (log2), or ratios between selection magnitudes estimated for DNase accessible regions to those for all promoter and intronic segments, the respective genomic control data is used in the both cases. Lower values of selection magnitude correspond to the stronger negative selection, thus negative fold change values correspond to stronger negative selection in DNase accessible regions. X axis displays different significantly conserved motifs (P < 0.05) for the set of promoter and intronic mutations. Data for affinity loss (A) and affinity gain (B) is presented for breast cancer (top subpanels) and lung adenocarcinoma (bottom subpanels). Members of FOX and NR transcription factor families are colored in blue and green.

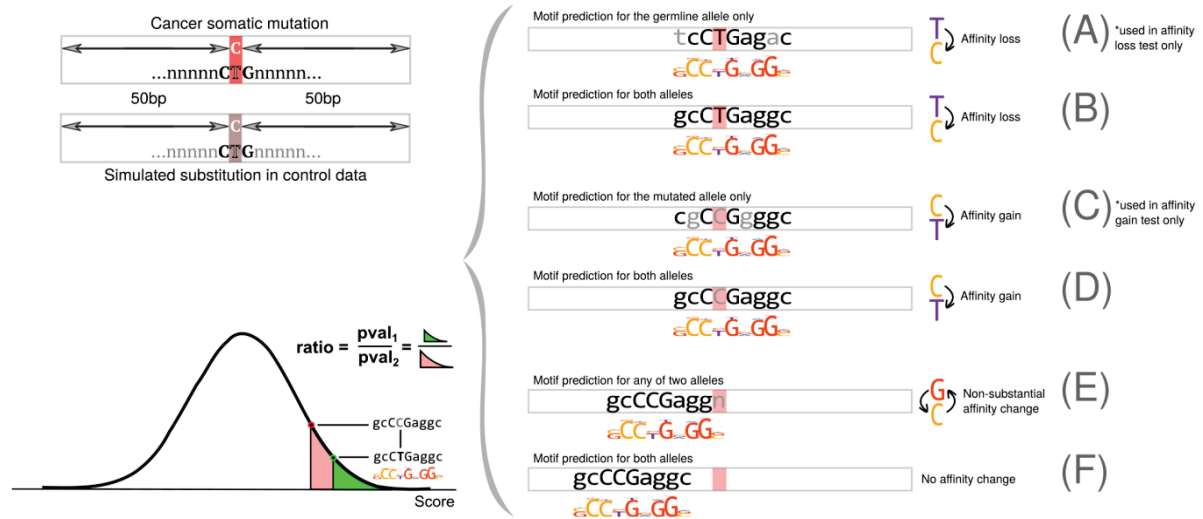

**Figure 4.** *Procedure used to evaluate affinity change events and estimate significance of difference between observed and expected frequencies.*

Top panel: prediction of binding sites in cancer and control data and evaluation of affinity change events. Bottom panel: binding sites predictions and affinity change events of AP-2 motifs; an example of 2x2 contingency table used to compute Fisher's exact test *P*-value.

**Table 1.** Examples of selection magnitude for conserved binding motifs and motifs frequently affected by somatic mutations. Selection magnitude estimated against two control data sets is given. Significant cases are colored by light green (protection from affinity change) and light red (frequent affinity change). Selected members of several transcription factor families are shown.

| Motif | Family | Breast | | | | Liver | | | | Lung Adeno | | | |
|---|---|---|---|---|---|---|---|---|---|---|---|---|---|
| | | Loss (vs genomic) | Loss (vs shuffle) | Gain (vs genomic) | Gain (vs shuffle) | Loss (vs genomic) | Loss (vs shuffle) | Gain (vs genomic) | Gain (vs shuffle) | Loss (vs genomic) | Loss (vs shuffle) | Gain (vs genomic) | Gain (vs shuffle) |
| AP2A | AP-2 | 1.218 | 1.208 | 0.911 | 0.909 | 1.063 | 0.928 | 1.114 | 0.855 | 1.082 | 1.011 | 1.175 | 1.091 |
| AP2B | AP-2 | 1.125 | 1.166 | 0.887 | 0.957 | 1.089 | 0.972 | 1.199 | 1.036 | 1.021 | 1.005 | 1.016 | 1.010 |
| CEBPA | C/EBP-related | 1.107 | 1.112 | 1.024 | 0.948 | 0.997 | 0.945 | 0.988 | 0.945 | 0.972 | 0.898 | 0.994 | 0.922 |
| CEBPB | C/EBP-related | 1.099 | 1.127 | 0.979 | 0.868 | 0.968 | 0.955 | 0.969 | 0.882 | 0.947 | 0.930 | 0.993 | 0.862 |
| CEBPE | C/EBP-related | 1.045 | 1.080 | 0.980 | 0.972 | 1.069 | 1.085 | 0.973 | 1.008 | 0.929 | 0.942 | 0.986 | 0.936 |
| CEBPG | C/EBP-related | 1.182 | 1.211 | 0.968 | 0.916 | 1.018 | 1.027 | 0.958 | 0.943 | 0.960 | 0.920 | 0.989 | 0.933 |
| EGR1 | Ets-related | 1.113 | 1.165 | 0.933 | 1.246 | 1.141 | 1.079 | 1.088 | 1.129 | 1.052 | 1.106 | 0.978 | 1.159 |
| ELF1 | Ets-related | 0.931 | 0.910 | 1.142 | 1.087 | 0.984 | 0.907 | 1.080 | 0.988 | 0.921 | 0.905 | 1.046 | 1.019 |
| ETS1 | Ets-related | 0.930 | 0.906 | 1.090 | 1.028 | 0.995 | 0.918 | 1.086 | 0.995 | 0.935 | 0.875 | 1.029 | 1.021 |
| FOXC1 | Forkhead box | 0.973 | 0.885 | 0.977 | 0.904 | 1.007 | 0.966 | 0.959 | 0.923 | 0.949 | 0.869 | 0.932 | 0.872 |
| FOXO1 | Forkhead box | 1.009 | 0.928 | 0.990 | 0.905 | 0.986 | 0.949 | 0.946 | 0.882 | 0.901 | 0.857 | 0.980 | 0.881 |
| FOXQ1 | Forkhead box | 0.978 | 0.904 | 0.942 | 0.922 | 0.993 | 0.978 | 0.927 | 0.955 | 1.002 | 0.930 | 0.960 | 0.956 |
| SP1 | Krüppel-related | 1.067 | 1.201 | 1.389 | 1.099 | 1.054 | 1.082 | 1.278 | 0.954 | 1.007 | 1.139 | 1.267 | 1.005 |
| SP2 | Krüppel-related | 0.996 | 1.112 | 1.132 | 1.158 | 1.038 | 0.980 | 1.114 | 1.041 | 0.949 | 1.017 | 1.083 | 1.083 |
| SP3 | Krüppel-related | 1.115 | 1.346 | 1.412 | 1.193 | 1.063 | 1.091 | 1.151 | 0.962 | 1.004 | 1.157 | 1.304 | 1.062 |
| ESR1_do | Steroid receptors | 0.838 | 0.688 | 0.815 | 0.636 | 1.047 | 0.860 | 1.017 | 0.819 | 0.9846 | 0.834 | 0.901 | 0.730 |
| ESR2_do | Steroid receptors | 0.880 | 0.746 | 0.852 | 0.609 | 1.033 | 0.866 | 0.999 | 0.743 | 1.0014 | 0.861 | 0.986 | 0.687 |

*Additional files are available at:*

*http://autosome.ru/supplement/addfil2015_3.zip*

**Additional file 1 (Format: PDF)**

**Supplementary Figure 1.** *Relative frequencies of non-coding mutation contexts in different cancer types.*

(top panel) Three cancer types with the largest number of mutation calls exhibit different mutation signatures. (bottom panel) Overall comparison of non-coding mutation signatures in 10 cancer types. The samples with lower numbers of total mutation calls display extreme contexts distributions. Cancer types are sorted by the total number of mutation calls. Mutations are grouped by the substitution (X > Y), the 5' and 3' nucleotides are shown in a lexicographical order.

**Additional file 2 (Format: PDF)**

**Supplementary Figure 2.** *Agreement between shuffle and genomic control data for detection of conserved motifs (negative selection) and motifs frequently altered (targeted) by mutations (positive selection).*

X axis shows the number of motifs, stacked bars show the number of significant motifs passing $P < 0.05$ for shuffle only (green), genomic only (light blue) and both (deep blue) controls. Data for three cancer types are shown. Panels: (A) Affinity loss. (B) Affinity gain.

**Additional file 3 (Format: XSLX)**

**Supplementary Table 1.** Mutation counts, frequencies of mutation contexts and relative control sizes for different cancer types. Initial as well as the final size (after binding sites predictions and equalizing the mutation contexts distribution) of each control set is shown.

**Additional file 4 (Format: XSLX)**

**Supplementary Table 2.** Raw data on binding sites predictions across mutation-centered windows in all evaluated cancer data sets as well as the respective generated control data sets.

**Additional file 5 (Format: XSLX)**

**Supplementary Table 3.** Binding motifs under positive and negative selection.
Data for separate motifs is shown, as well as aggregated data for the motifs grouped by family (number of significantly protected motifs is shown, family name and ID are given according to TFClass).

**Additional file 6 (Format: XSLX)**

**Supplementary Table 4.** Overview of DNase accessibility data analysis including mutation counts, motif predictions and data sources used to filter mutations in regulatory regions for breast cancer and lung adenocarcinoma.

**Additional file 7 (Format: XSLX)**

**Supplementary Table 5.** Binding motifs found under positive and negative selection in a basic test considering mutations overlapping and avoiding binding sites predictions. Raw data on binding sites predictions is given along with the resulting aggregated information for motif families.